\documentclass{article}
\begin{document}
\vbadness = 100000
\hbadness = 100000
\title{Lorentz contraction, Bell's spaceships, and rigid body motion in special relativity}
\author{Jerrold Franklin\footnote{Internet address:
Jerry.F@TEMPLE.EDU}\\
Department of Physics\\
Temple University, Philadelphia, PA 19122-6082}
\maketitle
\begin{abstract}
The meaning of  Lorentz contraction in special relativity and its connection with Bell's spaceships parable is discussed.  The motion of Bell's spaceships is then compared with the accelerated motion of a rigid body.
We have tried to write this in a simple form that could be used to correct students' misconceptions due to conflicting earlier treatments.

\end{abstract}

\section{`Lorentz contraction' in special relativity}

We have put the term  `Lorentz contraction' in quotes, because, as we will explain, Lorentz contraction is not what actually occurs for a moving object in special relativity (SR).  This is well known to most physicists, but too often `Lorentz contraction' is given a spurious physical reality.  Lorentz contraction is so called because 
\mbox{H. A. Lorentz} proposed an actual physical contraction of a moving object as an explanation of the null result of the Michaelson-Morley experiment, thus preserving the aether\cite{L}.  Lorentz's equation for the length $L'$ of a rod  moving at velocity $v$ was
\footnote{We are using units with $c=1$.}   
\begin{equation}
L'=L/\gamma,\quad{\rm with}\quad\gamma=\frac{1}{\sqrt{1-v^2}},
\label{eq:lc}
\end{equation}
where $L$ is the length the rod would have at rest.
The shrinking of a moving rod, with similar motivation, had been suggested previously by 
G. F. Fitzgerald\cite{f}, and the effect is often called `Lorentz-Fitzgerald contraction'.  

Lorentz and Fitzgerald attributed this physical contraction to new electromagnetic molecular forces within a moving rod with concomitant stresses and strains.  Although there is an equation identical to Eq.\ (\ref{eq:lc})  in special relativity, the Lorentz derivation and its application to the Michaelson-Morley experiment contradicts relativity.  Indeed, if the velocity of light is the same in all directions in any inertial frame, then applying the `Lorentz-Fitzgerald contraction' to the Michaelson-Morley apparatus would produce a positive result.

The equation for Lorentz contraction (We drop the quotes and Fitzgerald.) in SR is the same as originally given by Lorentz, but the physical significance is quite different.  {\bf In SR, Eq.\ (\ref{eq:lc}) relates the length  $L=|{\bf L\cdot{\hat v}}|$ of an object as measured in its rest system, to the length 
$L'=|{\bf L'\cdot{\hat v}}|$ of the same object as measured in a particular way in a system S$'$ that moves with constant velocity $\bf v$ with respect to the system S.} 

There are several important things to note in the wording of the previous sentence.
\begin{description}
\item(i) In order to compare the measured length of a moving object to its measured length in a system in which it is not moving, two different Lorentz systems, each with a constant velocity, are required.
\item(ii) In SR, there is no change in the object.  It is only the coordinate system that is changed from S to S$'$.
\item(iii) The measured length of a moving object depends on the `particular way' in which it is measured. 
\end{description}

The length of a stick which is at rest is measured by the difference $x_2-x_1$ of its ends.   For a stick at rest, the time at which each end is measured is unimportant, and they can be measured at different times.
In the usual textbook formulation of SR, the length of a moving stick is defined
by measuring the coordinates $x'_1$ and $x'_2$ of each end at the same time (simultaneously), $t'_1=t'_2$, 
in S$'$ and taking the length $L'$ as the difference
$x'_2-x'_1$.  This seems obvious in classical mechanics where it gives the same length to a moving stick as it has  at rest.  However, there are some difficulties and ambiguities associated with this definition, and there are other reasonable definitions of the length of a moving object.  For one thing, there is no a priori reason to expect a classical method of measurement to still be valid for relativistic motion.

An observer at rest in S will tell the measurers in S$'$, `Of course you got the wrong answer.  You measured each end at different times.  If you're moving past me, you should measure the distances at the same time.'  This argument occurs because, in SR, what is simultaneous in one Lorentz system is not simultaneous in others.   A witness in S would testify in court that the measurers in S$'$ made the length measurement incorrectly.  If the measurers in S$'$ made their measurements at each end when told to by those in S so that the measurement times were equal 
in S, their measured length $L'$ would be greater than $L$, not less.
In prerelativistic physics, it did not matter in which system the distance measurements were simultaneous, since in either case the measurement would give the same result.  But that is no longer true in relativity. 

It should also be pointed out that another classically reasonable method of measuring the length is to take a photograph of a moving object and compare it with a photograph of the same object at rest.  As Terrell\cite{t} showed some time ago, the photograph would show an object that is somewhat rotated, but of the same shape and dimensions as it had at rest.  Indeed, the photograph of a moving sphere would show a sphere of the same size.

These ambiguous definitions of `length' for a moving object arise from the fact that 
$x_2-x_1$ is not a Lorentz invariant, but only one component of a four-vector,
so the Lorentz transformed difference $x'_2-x'_1$  is just for this one component.
A Lorentz transformation between coordinate systems in relative motion is a generalized rotation in space-time.
Just as a three dimensional rotation changes the coordinate difference $x_2-x_1$ to $x'_2-x'_1$ , so does the four dimensional rotation in space-time.  And, just as the `shortening' of a stick that is rotated in three dimensions is an illusion, we now can see that the `shortening' of a stick that is rotated in four dimensions by a Lorentz transformation is also illusory.

This suggests the need for a definition of `length' that is the same for any state of uniform motion.  This would
correspond to the use in relativity of  `proper time' and `invariant mass' for time and mass, but the terms ``proper length' and `invariant length' have already been used in the literature with other meanings.     	       
The term we recommend for length is `rest frame length', which we define as the length a moving object has after a Lorentz transformation to its rest system.  If length is to be considered a physical attribute of an object, then this physical attribute should be the rest frame length. 
This length, of course, would not be changed by uniform motion.

One other point to be considered is whether strains and stresses can be induced by Lorentz contraction, as is contended in Refs.\ [1,2,4,5].  Our answer to this is clear from the previous discussion.  Just as a 3D rotation of an object does not induce strain, a 4D rotation (Lorentz transformation) will not induce strain and consequent stress.  We illustrate this with a simple example.  Consider a brittle wine glass at rest on a table.  If motion at constant velocity induced strain, then constant motion could shatter the wine glass.  However, by the first postulate of special relativity, moving the wine glass at constant velocity is equivalent to having the wine glass at rest and an observer moving past it at an equal, but opposite, velocity.  That means that just walking past (and looking at) a wine glass at rest on a table could shatter the wine glass.  We see from this example that it is only the rest frame length of an object that relates to strains or stresses on the object.  The process of accelerating an object to a constant velocity may induce strain, depending on how the acceleration  is applied, and we discuss this below in connection with the Bell spaceships.

To summarize this section:\\          
The belief that a moving object has a different length comes from using the prerelativistic notion that keeping 
$t'_2=t'_1$ permits correct length measurement of the moving object.  We see that this does not give an unambiguous length, and has sometimes incorrectly suggested that motion at constant velocity can induce strain.  We conclude that
the length of an object can be measured only in its rest system.  If the object is moving, making a Lorentz transformation to its rest system is the only way to get a reliable measurement of its length.  This is generally true of intrinsic properties of physical objects that are not Lorentz invariants. 
 
\section{The Bell spaceship paradox}

We first present the nexus of the Bell spaceship paradox as originally presented by John Bell\cite{bell}.  
Although Bell's name has been attached to the paradox, the thought experiment involved was first considered by 
Dewan and Beran\cite{db} as a demonstration `that relativistic contraction can introduce stress effects in a moving body.'  We have disputed this contention in the previous section. 
A large number of published\cite{papers} and unpublished papers of varying contentions and conclusions have been written in the years since Bell's original formulation.  While not addressing these subsequent papers here, we think our resolution of the paradox (as no paradox) applies to most of them.  A recent paper by Vesselin Petkov\cite{pet} comes to conclusions similar to ours about Bell's spaceships, and critiques a number of earlier papers.  

Bell considered two spaceships starting from rest in a Lorentz system S, and undergoing identical accelerations 
${\bf a}(t)$ in that system.  We analyze this situation now in some detail.  We denote the spaceships as L and R, each having acceleration $a(t)$ in the positive x direction in system S, starting from rest at positions $x_L=0$ for the nose of L  and $x_R=d$ for the tail of R, so the starting distance between the ships is $d$.  
At equal times in S with $t_L=t_R=t$, the spaceships will have equal velocities $v=v_L=v_R$, and the difference $x_R-x_L$ will remain constant at $d$.  

If, at a time when each spaceship has a velocity $v$, we make a Lorentz transformation with velocity $v$, each spaceship will be at rest and the distances $x'_L$ and $x'_R$ will be given by 
\begin{eqnarray}
x'_L&=&\gamma(x_L-vt)\\
x'_R&=&\gamma(x_L+d-vt).
\end{eqnarray}
The distance between the two ships in system S$'$ is given by
\begin{equation}
d'=x'_R-x'_L=\gamma d.
\label{eq:dpd}
\end{equation}
We see that this length is greater than the length between the spaceships before their motion.  That is, as the velocity in S increases, the distance between the spaceships in their rest system S$'$ increases.
Equation (\ref{eq:dpd}), as is usual for Lorentz contraction, states that the length between the moving ships, measured at equal times (for each ship),  is shorter than this length measured at different times in their common rest system.
The situation here is a bit different than in the usual discussion of Lorentz contraction, where the rest length is fixed and the measured moving length shortens.  Here, because of the way Bell constructed his example, the measured moving length is constant as the velocity increases.  Consequently, the rest frame length in the instantaneous rest system S$'$ must increase in accordance with Eq.\ (\ref{eq:dpd}), and get longer than the original rest frame length in system S.

Although the spaceships are accelerating, the system S$'$ is a Lorentz system moving at constant velocity.  Since each ship is instantaneously at rest in this system, the length 
$d'=\gamma d$ is the rest frame distance between the ships.  As such, it is the physical distance between the ships.  If there were an inextensible cable between the ships, it would snap at the start of motion of the ships.  An elastic cable would stretch until it reached its maximum possible length $d_{\rm Max}$, at which point it would snap.  That is, a cable connecting the two ships would snap when 
\begin{equation}
d'=\gamma d=d_{\rm Max}.
\end{equation}
The velocity at which the cable would snap is when
\begin{equation}
\gamma=1+(d_{\rm Max}-d)/d=1+s_{\rm Max},
\end{equation}
where $s_{\rm Max}$ is the maximum strain the cable can withstand.

Notice there is no hint of a paradox in the above treatment of two accelerating spaceships.  Bell's paradox was that his intuition told him the cable would break, yet there was no change in the distance between the ships in system S.  He suggested resolving the paradox by stating that a cable between the ships would shorten due to the contraction of a physical object proposed by Fitzgerald and Lorentz, while the distance between the ships would not change.  This resolution however contradicts special relativity which allows no such difference in any measurement of these two equal lengths.

A question might be raised by the fact that, although each spaceship is instantaneously at rest in system S$'$, the $x'_R$ and $x'_L$ measurements are made at different times.
Since the ships are accelerating, won't that affect the measurement?  
Without acceleration, it is clear that length measurements can be made in the rest system at different times for each ship.  When there is acceleration,
we first consider a case where the two ships are initially at rest with no acceleration.  Then the distance measurement for each ship can be made at any time.  If each ship starts to accelerate after its distance measurement is made, this can't affect a measurement which has already been made.  The ship's distance will be the same  immediately after acceleration is applied, because the velocity will still be negligible and the distance wouldn't have changed yet.  
A similar argument can be applied to a situation where the ships are originally accelerating and their acceleration is stopped at the instant each comes to rest in 
system S$'$.  Once the acceleration stops and the ship is at rest, it does not matter when each measurement is made.  The measurements could also be made just before the acceleration is turned off, because the velocity and distance moved will be negligible, as in the previous case.  We have come to the  reasonable conclusion that past events and future events cannot affect a distance measurement between ships, each of which was at rest at the time of measurement.  
This means that the ship's distances can be measured at different times in their common rest system even if each has acceleration.

A similar argument can be made in system S where each ship is moving with the same velocity and acceleration at the same time.  Future events and past events will not affect the distance between the ships, so the rest frame distance between them will be the same as if they had constant velocity at the time of measurement.  We conclude that the rest frame distance between the ships depends only on their common velocity and not on their acceleration.  The result is that when each spaceship has the same velocity $v$ at the same time, the rest frame distance between them is $d'=\gamma d$, even for continually accelerating spaceships . 

\section{Rigid body motion in special relativity}

In the motion described by Bell and in the preceding section, the acceleration of each spaceship is the same at equal times in system S.  This also corresponds to each having the same acceleration $a'$ in their instantaneous 
rest system\footnote{What we mean by `instantaneous rest system' is a Lorentz system moving at constant velocity in which the object is momentarily at rest.} if their rest system acceleration is constant in time.  
This is because their acceleration in system S, in which they each have velocity $v$, is given by\footnote{Equation 
(\ref{eq:aa}) follows from transformation equations for acceleration on page 569 of Ref.\ \cite{dj} and 
page 375 of Ref.\ \cite{jf} when $a'$ is the acceleration in the rest system.}
\begin{equation}
a=a'/\gamma^3.
\label{eq:aa}
\end{equation}
Thus, if the two spaceships have constant equal acceleration in their rest system S$'$, they will also have equal (but not constant) acceleration in system S.  This keeps their velocities the same, and the distance between the ships constant in system S.  But we have seen that the distance between the ships increases in their mutual rest system.

In the motion of a rigid body,  the dimensions of the object are unchanged by acceleration.  This type of motion in special relativity was first considered by Max Born and is often called `Born rigid motion'\cite{mb}, but our treatment of rigid body motion is somewhat different than Born's.
From the preceding paragraph, we see that keeping lengths constant in the rest system requires different rest frame accelerations for different parts of a rigid body.  The two spaceships are a simple example of this in that we have to consider separate accelerations for each ship.  To simplify the treatment, we consider the two accelerations $g_R$ and $g_L$ to be two different constants in time.

We first consider the motion of either spaceship in system S.
The acceleration there is related to the rest frame acceleration $g$ 
 by\footnote{Although we use the symbol $g$, it is not the acceleration due to gravity.}  
\begin{equation}
g=\gamma^3 a=\frac{d}{dt}(\gamma{\bf v}),
\end{equation}
where we have used the relation
\begin{equation}
\frac{d}{dt}(\gamma{\bf v})
= \frac{d}{dt}\left[\frac{\bf v}
{\sqrt{1-{\bf v}^2}}\right]
=\gamma{\bf a}+\gamma^3{\bf v(v\cdot a)}]
=\gamma^3{\bf a},
\label{eq:acc}
\end{equation}
for {\bf a} parallel to {\bf v}.
We solve this differential equation by the following steps:
\begin{eqnarray}
g&=&\gamma^3 a=\frac{d(\gamma v)}{dt}\nonumber\\
gt&=&\gamma v=\frac{v}{\sqrt{1-v^2}}\nonumber\\
v&=&\frac{gt}{\sqrt{1+g^2t^2}}=\frac{dx}{dt}\nonumber\\
\int_{x_0}^x dx'&=&\int_0^t\frac{g\bar{t}d\bar{t}}{\sqrt{1+g^2\bar{t}^2}}\nonumber\\
x&=&x_0+\left(\sqrt{1+g^2t^2}-1\right)/g.
\label{eq:deriv}
\end {eqnarray} 

For the two ships R and L, Eq.\ (\ref{eq:deriv}) gives
\begin{eqnarray}
x_R&=&d+\left(\sqrt{1+g_R^2t_R^2}-1\right)/g_R\\
x_L&=&\left(\sqrt{1+g_L^2t_L^2}-1\right)/g_L.
\end {eqnarray}
We see that if $g_L=g_R$, the distance between the two spaceships remains constant at $d$ for equal times in system S, in which they are each moving.

Rigid body motion for the two spaceships means keeping the distance between the ships constant at $d$ in their mutual rest system, either by appropriately adjusting their thrusts, or by connecting them with a cable of fixed length.  In order to transform to the mutual rest system of R and L, we have to know $x_R$ and $x_L$ when they have equal velocities in S.  We can do this by using the relations
\begin{equation}
t=\gamma v/g\quad{\rm and}\quad\gamma=\sqrt{1+g^2t^2},
\end{equation}
which follow from the steps in Eq.\ (\ref{eq:deriv}) above. 
 Then, we have
 \begin{eqnarray}
x_R&=&d+(\gamma-1)/g_R\\
x_L&=&(\gamma-1)/g_L
\end {eqnarray}    
for the position of each spaceship with the same velocity $v$.  Of course, the two times $t_R$ and $t_L$ are now different.  The times are given by
 \begin{eqnarray}
t_R&=&\gamma v/g_R\\
t_L&=&\gamma v/g_L.
\end{eqnarray} 

The condition that the distance between the ships in their rest system be fixed at $d$ can be imposed by Lorentz transforming the difference in the ship's positions in system S to their rest system.  
The space and time differences for the two spaceships are
 \begin{eqnarray}
\Delta x&=&d+(\gamma-1)\delta\\
\Delta t&=&\gamma v\delta,
\end{eqnarray} 
where 
\begin{equation}
\delta=\frac{1}{g_R}-\frac{1}{g_L}.
\end{equation}
The Lorentz transformation to the rest frame is
\begin{eqnarray}
d&=&\Delta x'=\gamma(\Delta x-v\Delta t)\nonumber\\
&=&\gamma[d+(\gamma-1)\delta-v^2\gamma\delta]\nonumber\\
&=&\gamma d+(1-\gamma)\delta,
\end{eqnarray}
so that
\begin{equation}
\delta=d=\frac{1}{g_R}-\frac{1}{g_L},
\end{equation}
and
\begin{equation}
g_L=\frac{g_R}{1-g_R d}.
\label{eq:gl}
\end{equation}

Thus there is a fixed relation between the constant accelerations of the two spaceships in their instantaneous rest system.  Maintaining these different rest frame accelerations for each ship will keep the rest frame distance between them constant.  For a rigid body, the rest frame acceleration throughout the body would be given  by $g_L$ in 
Eq.\ (\ref{eq:gl}) with $g_R$ being the acceleration of the front end and $d$ the $x$ distance from the front end.  
We see that in order to keep the body rigid in its rest frame, the acceleration has to vary. 
Although the acceleration varies throughout the rigid body, there will be no strain because this varying acceleration preserves the rest frame dimensions of the body.  Any stress in the body, will not be appreciably different than the stress induced by nonrelativistic acceleration of a rigid body.  This is to be contrasted with the case of a cable between the two Bell spaceships, which will have ever increasing strain from the beginning of the motion.

For rigid body motion, the time difference in the rest frame is given by the Lorentz transformation
\begin{eqnarray}
\Delta t'&=&\gamma(\Delta t-v\Delta x)\nonumber\\
&=&\gamma[\gamma v\delta-vd-v(\gamma-1)\delta]\nonumber\\
&=&\gamma v\delta-\gamma vd=0.
\end{eqnarray}
Thus, the two rest frame times $t'_R$ and $t'_L$ are equal and there is no question that $d$ remains the constant rest frame distance between the ships.

\section{Conclusion}

We have seen that the physical length of an object is the rest frame length as measured in the instantaneous rest frame of the object.  For two spaceships having equal accelerations, as in Bell's spaceship example,
the distance between the moving ships appears to be constant, but the rest frame distance between them continually increases.  This means that a cable between the two ships must eventually break if the acceleration continues.  For rigid body motion in special relativity, the rest frame acceleration throughout the body will vary as in Eq.\ (\ref{eq:gl}), but this will preserve the rest frame dimensions of the object.  
We have only treated rigid body motion with constant acceleration in the instantaneous rest frame.  For a general time dependence of acceleration, our Bell spaceship discussion would not be affected, but the spatial dependence of the rest frame acceleration for a rigid body would be more complicated than  that in Eq.\ (\ref{eq:gl}).

\end{document}